\documentclass[conference]{IEEEtran}
\IEEEoverridecommandlockouts
\usepackage{cite}
\usepackage{amsmath,amssymb,amsfonts}
\usepackage{graphicx}
\usepackage{tikz}
\usepackage[table]{xcolor}
\usepackage{booktabs}
\usepackage{array}
\usepackage{multirow}
\usepackage{float}
\usepackage{hyperref}
\usepackage{enumitem}
\usetikzlibrary{positioning,arrows.meta,shapes.geometric,fit,backgrounds}
\hypersetup{colorlinks=true,linkcolor=blue,urlcolor=blue,citecolor=blue}

\definecolor{cellfilled}{RGB}{255, 230, 200}
\definecolor{cellempty}{RGB}{235, 245, 235}
\definecolor{pipeblue}{RGB}{137, 207, 240}
\definecolor{pipegrey}{RGB}{220, 225, 232}

\begin{document}

\title{An End-to-End Threat Model for the Quantum-as-a-Service Pipeline \\

\thanks{This work has been supported by the Business Finland through project \textit{SeQuSoS} (Grant No. 112/31/2024), and Research Council of Finland through project \textit{Profi 8 - qSIME} (Grants No. 365343).}
}

\author{\IEEEauthorblockN{Badhon Rahman}
\IEEEauthorblockA{\textit{Faculty of Information Technology} \\
\textit{University of Jyväskylä}\\
Jyväskylä, Finland \\
badhon.b.rahman@jyu.fi}
\and
\IEEEauthorblockN{Majid Haghparast}
\IEEEauthorblockA{\textit{Faculty of Information Technology} \\
\textit{University of Jyväskylä}\\
Jyväskylä, Finland \\
majid.m.haghparast@jyu.fi}
\and
\IEEEauthorblockN{Tommi Mikkonen}
\IEEEauthorblockA{\textit{Faculty of Information Technology} \\
\textit{University of Jyväskylä}\\
Jyväskylä, Finland \\
tommi.j.mikkonen@jyu.fi}

}

\maketitle

\begin{abstract}
Cloud-based accessing of Quantum-as-a-Service (QaaS) platforms such as IBM Quantum, IonQ Cloud, and Amazon Braket is becoming popular day by day. Hybrid quantum-classical algorithms (VQE, QAOA, QML) transfer data via a long layered pipeline of orchestration, compilation, and execution. Recent works have demonstrated various critical attacks at individual stages: Calibration tampering, SWAP attacks, QubitHammer, and so on. However, these attacks remain separated because of their own terminology, and existing STRIDE-based threat modeling in the context of quantum lacks a structured view towards the QaaS stack itself. We address this concern by decomposing the workflow into six-stage model with STRIDE threat modeling. Our matrix demonstrated attack vectors in quantum-specific, inherited classical, and plausible tiers for each of the stages. We further investigate the underexplored sections (repudiation and elevation-of-privilege) and distinguish three different cross-stage attack chains with higher impacts.  
\end{abstract}


\section{Introduction}\label{sec:sec1}
Quantum computers are increasingly being used as a cloud service. Platforms like IBM Quantum, Amazon Braket, and IonQ Cloud expose the quantum processors behind classical APIs. The current working process of quantum computation is also hybrid. Now, a client designs and simulates their circuits from a classical device, transmits towards a quantum hardware as a job, after execution delivers the measurement output and updates the parameters. Hybrid algorithms such as VQE, QAOA, and QML follow the same process pattern. This hybrid architecture introduces various novel attack exposures across orchestration layers and the classical-quantum interface \cite{b11}. 

This complete pipeline is a rich attack surface for critical exposures. Recent research work has demonstrated diverse threats in this multi-tenant environment. Such as SWAP attack \cite{b9}, QubitHammer \cite{b10}, Inverse-transpilation \cite{b12}, parameter and measurement tampering by untrustworthy vendors \cite{b17},  calibration tampering via DoS \cite{b7}, pulse-level abuse, and many more. 

However, these attacks remain siloed. Each of these attacks is demonstrated under its own terminology. For instance, Side-channel and Fault-injection attacks cannot be placed under a common map. Therefore, in this poster, we presented a six-stage STRIDE threat model for the long QaaS pipeline showcases attack vectors for each stage and investigated underexplored sections. A unified attack-surface map of the QaaS pipeline is shown as a result.
\section{Background and Related Work}\label{sec:sec2}
\textbf{STRIDE} \cite{b5}\cite{b6} threat model break downs threats into \emph{Spoofing, Tampering, Repudiation, Information disclosure, Denial of service, and Elevation of privilege}.

In prior threat modeling work, Lee and Tan \cite{b2} employ STRIDE/PASTA to the universal Cyber Physical System (CPS) that is exposed to quantum cryptanalysis. Baseri et al. \cite{b3} applied STRIDE for the quantum-induced attacks on classical cryptography. Neither of these focuses on the quantum cloud service itself. Furthermore, recent studies Swaroop et al. \cite{b7}, Samuel et al. \cite{b4}, and Justin et al. \cite{b8} have listed attack vectors within the hardware and environment, but without a unified threat model. Our prior work \cite{b1} classified Quantum Software Engineering (QSE) security challenges has identified this gap; this study operationalizes and provides structured threat modeling. Eventually, no previous study with a unified service-level threat model was found for the quantum cloud stack.

\section{Threat Modeling}\label{sec:sec3}
We have designed and divided our hybrid quantum-classical model into 6 stages: \textbf{S1 - Developer Environment}: Local circuit simulation with an SDK (Qiskit, Cirq, Q\#, Qrisp). \textbf{S2 - Authentication \& Submission}: API token, SDK calls, IAM (Identity and Access Management). \textbf{S3 - Cloud Orchestration \& Compilation}: Queue, scheduling, transpiler. \textbf{S4 - Quantum Hardware Execution}: QPU with calibration of metadata, pulse-level control, in multi-tenant. \textbf{S5 - Result Return Path}: Measurements, error mitigation, and post-processing. \textbf{S6 - Hybrid Iteration Loop}: Repeating the workflow from S2 to S5.

\section{Results}\label{sec:sec4}
Table \ref{tab:matrix} depicts the STRIDE-based threat modeling with our six different defined stages of the hybrid quantum-classical model. 

\begin{table*}[!t]
\centering
\caption{STRIDE-based threat modeling with Pipeline-Stage Threat Matrix. Orange cells indicate published attacks, light blue cells are inherited classical attack vectors, and light green cells indicate plausible but under-studied threats.}
\label{tab:matrix}
\renewcommand{\arraystretch}{1.05}
\setlength{\tabcolsep}{2.5pt}
\scriptsize
\begin{tabular}{|>{\columncolor{pipegrey}}p{1.3cm}|p{2.45cm}|p{2.65cm}|p{2.4cm}|p{2.75cm}|p{2.6cm}|p{2.3cm}|}
\hline
\rowcolor{pipegrey}
& \textbf{Spoofing} & \textbf{Tampering} & \textbf{Repudiation} & \textbf{Info.\ disclosure} & \textbf{Denial of service} & \textbf{Elev.\ of priv.} \\
\textbf{S1 Dev.}
& \cellcolor{pipeblue}Compromised SDK / typosquatted dependency
& \cellcolor{pipeblue}Malicious SDK extension / IDE plugin
& \cellcolor{cellempty}-- 
& \cellcolor{pipeblue}Local circuit / key leak
& \cellcolor{cellempty}Local exhaustion
& \cellcolor{cellempty}-- \\
\hline
\textbf{S2 Auth}
& \cellcolor{pipeblue}Token theft, IAM impersonation
& \cellcolor{pipeblue}In-flight circuit modification
& \cellcolor{cellempty}Weak QPU job audit
& \cellcolor{pipeblue}Network eavesdropping
& \cellcolor{pipeblue}Quota / queue flood
& \cellcolor{cellempty}IAM role escalation \\
\hline
\textbf{S3 Compile}
& \cellcolor{cellempty}Compiler service spoof
& \cellcolor{cellfilled}Untrusted compiler injects \cite{b7}
& \cellcolor{cellempty}-- 
& \cellcolor{cellfilled}Inverse-transpile \cite{b12}; QML theft \cite{b16}
& \cellcolor{cellfilled}Scheduler abuse, transpile starvation \cite{b7}
& \cellcolor{cellempty}Malicious compiler pass \\
\hline
\textbf{S4 QPU}
& \cellcolor{cellempty}--
& \cellcolor{cellfilled}SWAP Attack \cite{b9}; QubitHammer \cite{b10}; Pulse abuse \cite{b14}
& \cellcolor{cellempty}No verifiable execution log
& \cellcolor{cellfilled}Passive SWAP \cite{b9}; Power Side-Channel\ \cite{b13}
& \cellcolor{cellfilled}Calibration tampering  \cite{b7}
& \cellcolor{cellfilled}Pulse-Level access abuse\cite{b14} \\
\hline
\textbf{S5 Return}
& \cellcolor{cellempty}Result-source spoof
& \cellcolor{cellfilled}Measurement tampering \cite{b17}
& \cellcolor{cellempty}No result chain
& \cellcolor{cellempty}Output stats leak via post-processing
& \cellcolor{cellempty}Channel flood
& \cellcolor{cellempty}-- \\
\hline
\textbf{S6 Loop}
& \cellcolor{cellempty}Spoofed optimizer endpoint
& \cellcolor{cellfilled}Iterative param tamper \cite{b17}
& \cellcolor{cellempty}Per-iter audit gap
& \cellcolor{cellfilled}Trained-parameter extraction \cite{b16}
& \cellcolor{cellfilled}Convergence stall \cite{b17}
& \cellcolor{cellempty}-- \\
\hline
\end{tabular}
\end{table*}

Throughout the stages, three Cross-stage attack chains are available and expose one valuable question: \textit{which stage combinations compose into higher impact attacks?}
\textbf{Chain~A (Side-channel identification $\to$ Targeted crosstalk):} Passive SWAP or controller power side-channel attacks at S4 reveal victim's circuit fingerprint (ansatz, gate information), and an adversary can implement QubitHammer/SWAP attack.
\textbf{Chain~B (Calibration topology $\to$ Targeted pulse placement):} Publicly exposed information (coupling map / Data) at S4, lets an adversary plan and place qubits, combining with on-device frequency sweep, supplying the parameters QubitHammer needed without observing victim's circuit. 
\textbf{Chain~C (Compiler IP leak $\to$ Transpile-stable Trojan):} Inverse-tranpilation at S3 discloses the compiler's optimization information, letting an adversary craft Trojan insertions that bypass known optimizations and turn into active S3 tampering. 

\section{Conclusion}\label{sec:sec5}

Our work presented a six-stage STRIDE threat model for the QaaS pipeline, demonstrated an organized quantum-specific, inherited classical, and plausible attack vectors into a single structure matrix. We further analyzed the underexplored section (repudiation and elevation-of-privilege) and characterized three cross-stage attack chains that compose stage local capabilities into higher impact threats.

\end{document}